\documentclass[10pt]{extarticle}

\usepackage[bordercolor=white,backgroundcolor=gray!30,linecolor=black,colorinlistoftodos, textsize=10pt]{todonotes}

\usepackage{amssymb}

\usepackage{textcomp}
\usepackage{amsmath}
\usepackage[textstyle,amssymb]{SIunits}
\usepackage{graphicx}
\usepackage{fixltx2e}  %for textsubscript
\usepackage[normalem]{ulem}  % for underline allowing linebreaks

\usepackage[twocolumn,textwidth=183mm,columnsep=6mm,top=20mm,bottom=25mm]{geometry}
\usepackage{helvet}
\usepackage{titlesec}
\titleformat*{\section}{\bfseries\sffamily}
\titlespacing{\section}{0pt}{*4}{*0}
\titleformat{\subsection}[runin]{\normalfont\bfseries}{\thesubsection.}{3pt}{}
\usepackage{setspace}
\usepackage[breaklinks=true]{hyperref} %[breaklinks=true]

\usepackage[font={footnotesize,sf},labelfont={sf,bf},labelsep=endash,justification=raggedright]{caption}

\usepackage{natbib}
\usepackage[labelfont=bf,justification=justified]{caption}

\begin{document}
%\keywords{lab-on-a-chip, quantum cascade laser, quantum cascade detector, surface plasmon polariton, dielectric loaded, monolithic}

\twocolumn[\begin{@twocolumnfalse}
	%******************* title *******************
	{\huge\sf \textbf{Coherent injection locking of quantum cascade laser frequency combs}}
	\vspace{0.5cm}
	
	%******************* authors *******************
	{\sf\large \textbf {Johannes~Hillbrand$^1$, Aaron~Maxwell~Andrews$^{1,2}$, Hermann~Detz$^{1,3}$, Gottfried Strasser$^{1,2}$, Benedikt~Schwarz$^{1,*}$}}
		\vspace{0.5cm}
		
		{\sf \textbf{$^1$Institute of Solid State Electronics, TU Wien, Floragasse 7, 1040 Vienna, Austria\\
				$^2$Center for Micro- and Nanostructures, TU Wien, Floragasse 7, 1040 Vienna, Austria\\
				$^3$CEITEC, Brno University of Technology, Brno, Czech Republic\\
				$^*$e-mail: {benedikt.schwarz@tuwien.ac.at}}}
		\vspace{0.5cm}
\end{@twocolumnfalse}]
\vspace{0.5cm}

%******************* abstract *******************
{\sf \small \textbf{\boldmath
		\noindent Quantum cascade laser (QCL) frequency combs are a promising candidate for chemical sensing and biomedical diagnostics, requiring only milliseconds of acquisition time to record absorption spectra without any moving parts~\cite{faist1994quantum,hugi2012mid,geiser2018single,villares2014dual}. 
		They are electrically pumped and have a small footprint, making them an ideal platform for on-chip integration~\cite{villares2015chip}. Until now, optical feedback is fatal for frequency comb generation in QCLs and destroys intermodal coherence~\cite{jouy2017dual}. This property imposes strict limits on the possible degree of integration. 
		Here, we demonstrate coherent injection locking of the repetition frequency to a stabilized RF oscillator.		 
		For the first time, we prove that the spectrum of the injection locked QCL can be phase-locked, resulting in the generation of a frequency comb. We show that injection locking is not only a versatile tool for all-electrical frequency stabilization, but also mitigates the fatal effect of optical feedback on the frequency comb. A prototype self-detected dual-comb setup consisting only of an injection locked dual-comb chip, a lens and a mirror demonstrates the enormous potential for on-chip dual-comb spectroscopy. These results pave the way to miniaturized and all-solid-state mid-infrared spectrometers.
	}
}

Optical frequency combs are lasers whose spectrum consists of a multitude of equidistant lines~\cite{udem2002optical}. They have emerged as high-precision tool for time metrology, frequency synthesis and spectroscopy~\cite{udem1999absolute,holzwarth2000optical,bernhardt2009cavity}. The mid-infrared (MIR) region is of particular interest for spectroscopic applications because most molecules exhibit fundamental roto-vibrational absorption lines in this portion of the electro-magnetic spectrum~\cite{haas2016advances,waclawek2014quartz}. In the last decade, the quantum cascade laser~\cite{faist1994quantum} (QCL) has become the dominant source of coherent MIR light.
Thanks to the large third-order optical non-linearity of their active regions, the longitudinal cavity modes of Fabry-P\'{e}rot QCLs can be locked to each other by four-wave-mixing (FWM) resulting in the generation of phase-coherent frequency combs~\cite{hugi2012mid,friedli2013four}. The equidistant teeth of the comb beat together causing a modulation of the laser intensity at the cavity roundtrip frequency. This beatnote can be measured using a photodetector that is fast enough to follow the beating. In the comb regime, all cavity modes are equidistant, which results in a narrow and stable beatnote with linewidth on the Hertz-level.
Depending on the laser bias, a second regime called high phase-noise regime is observed, where increased amplitude and phase-noise of the comb lines result in a considerably broader beatnote. This regime is caused by the finite laser dispersion\cite{faist2016quantum} and limits the applicability of the combs for dual-comb spectroscopy~\cite{villares2014dual}.
Due to the fast carrier dynamics in QCL active regions, the beating of the comb lines induces a temporal modulation of the population inversion.
This population pulsation results in a macroscopic current modulation, enabling the direct observation of the so called electrical beatnote.

\begin{figure}[b!]
	\centering
	\includegraphics[width=0.90\linewidth,trim={0 0.45cm 0 0},clip]{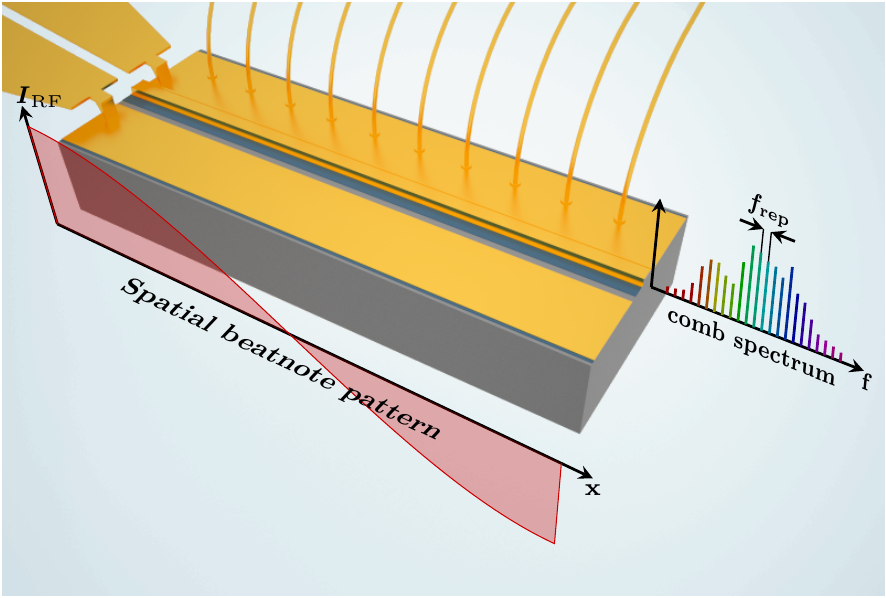}
	\caption{\textbf{Schematic of the device.} Injection locking of the coherent beating in a QCL frequency comb using an RF probe landed ontop of the edge of the laser cavity. The spatial pattern of the electrical beatnote is highlighted in red. Due to the standing waves in the Fabry-P\'{e}rot cavity, the amplitude of the beatnote current $I$ behaves according to $I\propto\cos \left(\pi x/L \right)$, where $L$ is the cavity length.}
	\label{fig:fig1}
\end{figure}

\begin{figure*}
	\centering
	\includegraphics[width=0.98\linewidth]{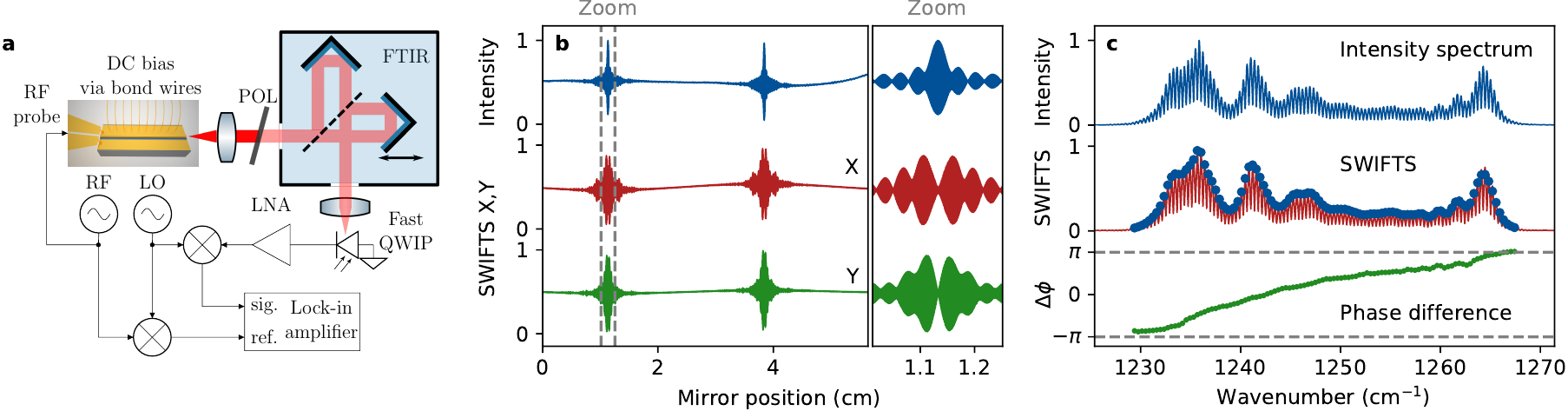}
	\caption{\textbf{SWIFTS analysis of the free-running comb.} \textbf{a}: SWIFTS setup. The optical beatnote of the QCL is detected by the QWIP through an FTIR, amplified by a low-noise amplifier (LNA) and mixed down to approximately 40 MHz using a local oscillator (LO). The mixing product of the LO and the oscillator used for injection locking (RF) acts as reference for the lock-in amplifier. A polarizer (POL) is used for adjustable attenuation. \textbf{b}: Intensity and SWIFTS quadrature interferograms of the free-running QCL frequency comb with zoom around zero path. \textbf{c}: Intensity spectrum (blue line), SWIFTS spectrum (red line) with expected SWIFTS amplitudes for full coherence (blue dots), as well as the phase differences between adjacent comb lines retrieved from the SWIFTS data (green dots). The corresponding reconstructed time domain signal can be found in supp. Fig. 1.}
	\label{fig:fig2}
\end{figure*}

As a general property of any oscillator, its frequency and phase can be locked to another oscillator provided that there is enough coupling~\cite{razavi2004study}. Consequently, an external RF modulation close to the roundtrip frequency injected into the QCL should be able to influence or even lock the electrical beatnote.
This concept is called electrical injection locking and was demonstrated by embedding a Fabry-P\'{e}rot QCL into a microstrip RF waveguide~\cite{stjean2014injection}. While this result proves that the collective action of the beatings of adjacent comb lines can be locked, it remains unclear to what extent the single beatings are locked and what their phases are. Indeed, another experiment revealed that the narrow beatnote of an injection locked QCL was produced by only a few modes, whereas the rest of the spectrum remained unlocked\cite{hugi2012mid}.

In this letter, we provide conclusive proof that all teeth of the comb can be locked coherently to an external RF source by applying the RF signal only to the end of the laser cavity.
This is the section of the cavity, where the electrical beating is most susceptible to the injected signal due to its inherent spatio-temporal pattern that was unveiled by recent work~\cite{piccardo2018time}. A schematic is shown in Fig~\ref{fig:fig1}.

Proving frequency comb operation of a QCL is a challenging task because the fast gain recovery time prevents the formation of short and intense pulses. Consequently, traditional methods based on non-linear autocorrelation techniques~\cite{kane1993characterization,iaconis1998spectral} cannot be employed. Instead, we use a linear phase-sensitive autocorrelation method called '\textit{Shifted Wave Interference Fourier Transform Spectroscopy}'\cite{burghoff2014terahertz,burghoff2015evaluating} (SWIFTS, Fig. \ref{fig:fig2}a). SWIFTS enables the direct measurement of the coherence and phases of the emitted comb spectrum.
The light emitted by the QCL is shined through a Fourier transform infrared (FTIR) spectrometer and detected by a fast photodetector. For this purpose, we designed and fabricated an RF optimized quantum well infrared photodetector (QWIP)\cite{schneider06quantum} matched to the laser emission wavelength.
By measuring the two quadratures $X$ and $Y$ of the optical beatnote as function of the mirror delay $\tau$ we obtain the complex interferogram of the portion of light that is locked to the RF oscillator. Contributions of mode pairs that are beating at another frequency are filtered by the lock-in. Subsequently, we retrieve the SWIFTS spectrum by applying a fast Fourier transform to the complex interferogram. In order to discuss this in more detail, we consider the electric field of the comb that is composed of discrete modes with amplitudes $A_n$ and frequencies $\omega_n=\omega_0+n \omega_{r}$, where $n$ is an integer and $\omega_0$ and $\omega_{r}$ are the carrier envelope offset frequency and repetition frequency of the comb. The complex SWIFTS spectrum is then given by
\begin{align}
\mathcal{F} \left(X+i Y \right)(\omega)=\sum_n & |A_n| |A_{n-1}|e^{i(\phi_n-\phi_{n-1})} \label{SWIFTSspectrum}\\&\times\biggl[ \delta \left(\omega-\omega_{n-\frac{1}{2}}\right)+\delta\left(\omega-\omega_r/2\right) \biggr]. \nonumber
\end{align}

\begin{figure*}[!t]
	\centering
	\includegraphics[width=0.98\linewidth]{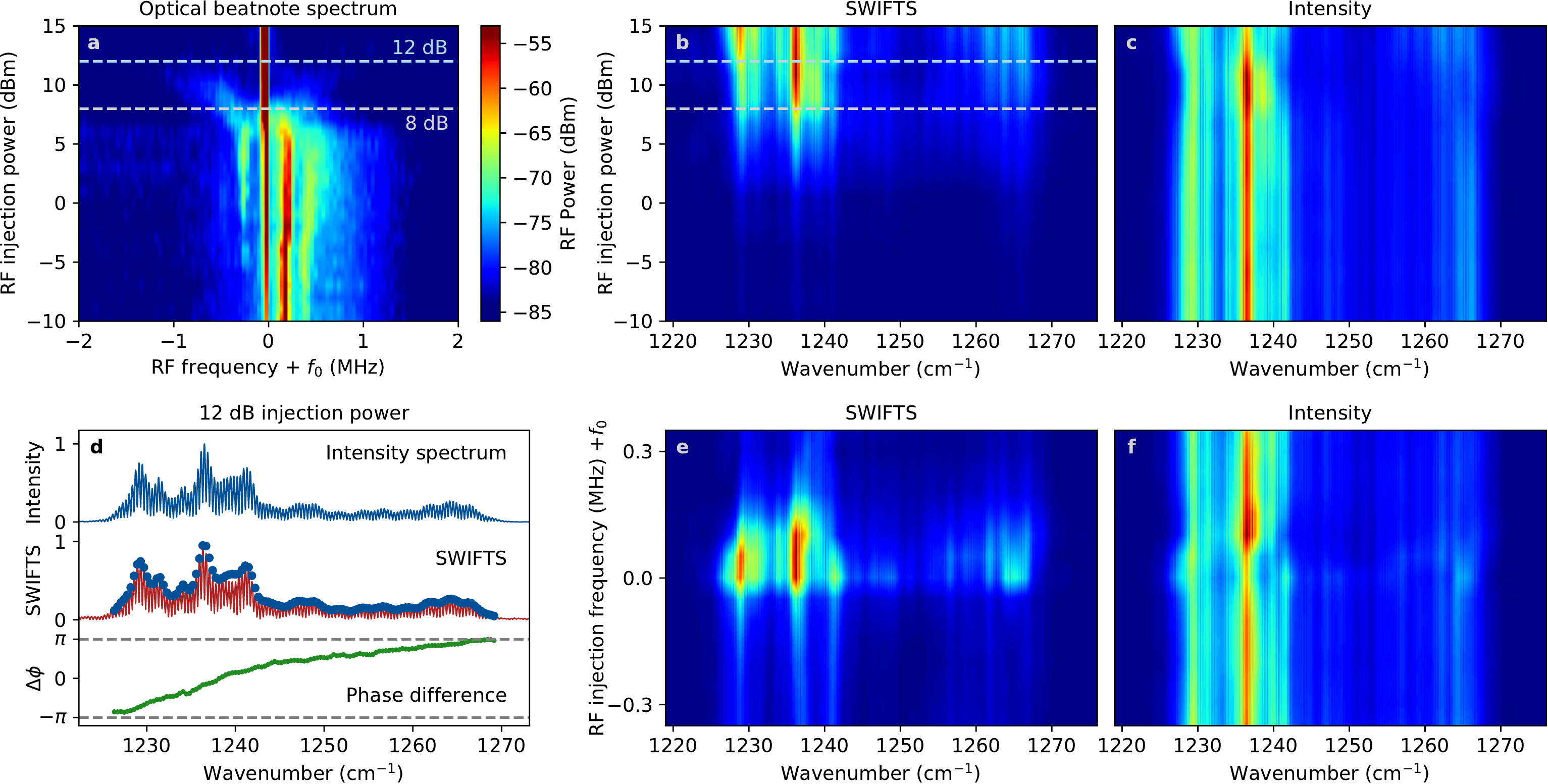}
	\caption{\textbf{Coherent injection locking.} \textbf{a}: RF spectra of the optical beatnote depending on the RF injection power recorded with the QWIP. \textbf{b}: SWIFTS spectra as function of injected power. \textbf{c}: Intensity spectra corresponding to (b). \textbf{d}: Detailed SWIFTS characterization at 12$\,$dBm. \textbf{e}: SWIFTS spectra for different injection frequencies at 12$\,$dBm injected power. The corresponding optical beatnote spectrum is shown in supplementary Fig. 3. \textbf{f}: Intensity spectra corresponding to (\textbf{e}).  }
	\label{fig:fig3}
\end{figure*}
The complex SWIFTS spectrum in eq.~\ref{SWIFTSspectrum} contains the phase difference of adjacent comb lines. Since only the part of the light locked to the phase reference is measured by the lock-in, the SWIFTS amplitudes are a direct and spectrally resolved measure of the intermodal coherence. If two modes are fully coherent (i.e., phase-locked), the SWIFTS amplitude is commensurate with the geometric average of the amplitudes $|A_n A_{n-1}|$ of the intensity spectrum. If, however, the relative phase-noise of a mode pair is non-zero, the SWIFTS amplitude decreases due to the narrow filter bandwidth of the lock-in.

Before investigating coherently injection locked fre\-quen\-cy combs, it is insightful to analyze the behavior the free-running comb, i.e. without RF injection or any other stabilization method. For this purpose, the electrical beatnote is used as phase reference.
A particularly interesting feature is already conspicuous in the recorded interferograms (see zoom in Fig. \ref{fig:fig2}b). Both SWIFTS quadratures have a minimum at zero-path difference of the FTIR mirrors, whereas the intensity interferogram has its maximum there. This phenomenon is related to the naturally favored comb state in QCLs, where the phases are arranged in a way that minimizes the amplitude of the optical beatnote and thus the amplitude modulation of the laser intensity.
This is due to the short sub-ps upper state lifetime in QCL active regions\cite{hugi2012mid, khurgin2014coherent}.
The corresponding SWIFTS spectrum (Fig. \ref{fig:fig2}c) has the same shape as the intensity spectrum without showing any spectral holes. This proves that indeed all teeth of the comb are phase-locked. The SWIFTS phases, i.e. the phase difference of adjacent comb lines, cover a range of $2\pi$ from the lowest to the highest frequency mode. This is consistent with the observation of a dominantly frequency modulated output of the QCL with a linearly chirped instantaneous frequency~\cite{singleton2018evidence}.

\begin{figure*}
	\centering
	\includegraphics[width=0.98\linewidth]{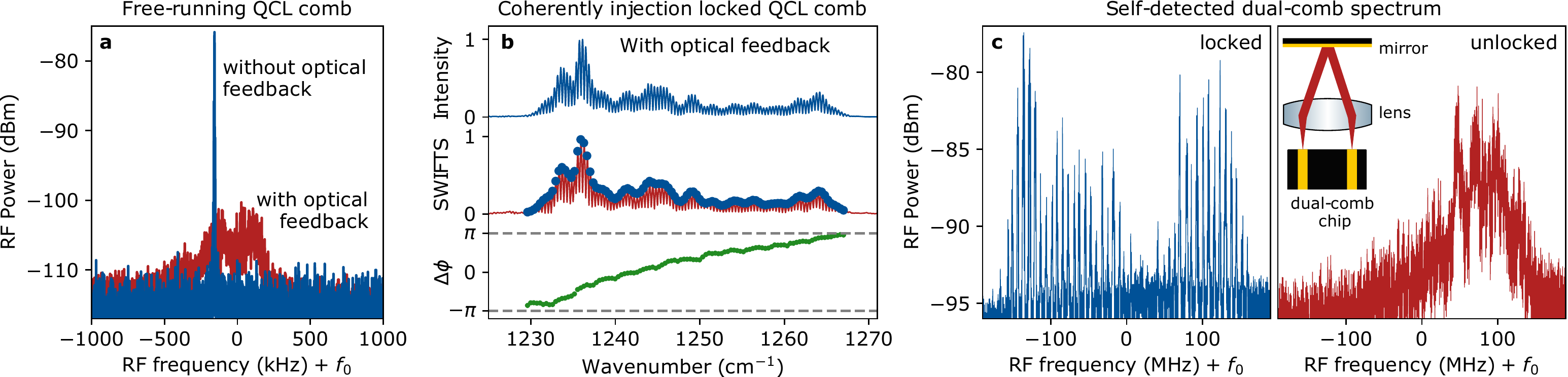}
	\caption{\textbf{Comb operation under strong optical feedback} \textbf{a}: Electrical beatnote of the free-running QCL without (blue line) and with (red line) strong optical feedback. \textbf{b}: SWIFTS characterization of the injection locked QCL in the presence of intense optical feedback. The periodic modulation of the optical spectrum and the phases is caused by Fabry-P\'{e}rot resonances and the Gires-Tournois effect in the Si wafer. The SWIFTS analysis in the same state without optical feedback is shown in supplementary Fig. 5. \textbf{c}: Self-detected multiheterodyne beat spectrum at $f_0=8.7\:$GHz of two QCLs that shine directly into each other (inset). The dual-comb spectrum (blue line) changes to a broad multiheterodyne beat (red line) when the injection frequency of one laser is detuned out of the locking range by 200 kHz.}
	\label{fig:fig4}
\end{figure*}

The first challenge to prove coherent injection locking is to show the capability to lock the QCL beatnote to the external oscillator. To do so, we drive the QCL at a bias where it operates in the high-phase noise regime (supplementary Fig. 2) and shine it directly on the fast QWIP. We then record the RF spectrum of the QWIP current with a spectrum analyzer while keeping the injection frequency fixed and $\approx$100~kHz below the beatnote (Fig. \ref{fig:fig3}a). As the injected RF power is increased, two sidebands appear above and below the frequency of the optical beatnote due to the mixing of the beatnote and the injected signal. The broad beatnote is pulled towards the frequency of the injected signal as the RF power is further increased to 5$\,$dBm and finally locks at 8$\,$dBm. Two sidepeaks that are roughly 20 dB weaker than the initial beatnote remain. At 12$\,$dBm, also these sidepeaks vanish and the microwave spectrum of the optical beating is fully controlled by the injected signal. The noise floor around the locked narrow beatnote is roughly 30 dB weaker than the peak power of the originally broad beatnote. This proves that the vast majority of the optical beatnote power is locked.
In order to highlight spectral regions which are locked to the external oscillator, we measure the SWIFTS and intensity spectrum as function of the injected power (Fig. \ref{fig:fig3}b and \ref{fig:fig3}c). The SWIFTS amplitude starts to grow considerably at 5$\,$dBm - equal to the power level at which the broad beatnote in Fig. \ref{fig:fig3}a is pulled towards the injection frequency. In the region between 8 and 12$\,$dBm, especially the SWIFTS amplitude of the comb lines around 1230$\:$cm$^{-1}$ and 1265$\:$cm$^{-1}$ increases suggesting that these modes are responsible for the weak sidepeaks of the optical beatnote in Fig. \ref{fig:fig3}a. A detailed snapshot of the SWIFTS characterization at 12$\,$dBm RF power (Fig. \ref{fig:fig3}d) shows that the SWIFTS amplitudes are commensurate with the values expected from the intensity spectrum. This proves that the entire spectrum of the QCL is phase locked to the RF oscillator. The SWIFTS phases feature the same phase pattern as observed in free-running comb operation, covering a range of $2\pi$ (Fig. \ref{fig:fig2}c). It is remarkable that the frequencies of the intermode beatings are locked to the external modulation while their phases remain in the natural state of a free-running QCL frequency comb instead of synchronizing to the injected signal. In order to investigate the influence of the injection frequency on the coherence of the QCL, we sweep it across the broad beatnote. The SWIFTS spectra (Fig. \ref{fig:fig3}e) show that the QCL is coherently locked to the injected signal in a narrow range of approximately 100 kHz around the frequency of the beatnote. Outside of this locking range, only a few strong modes contribute to the SWIFTS spectrum. The narrow locking range could explain why previous studies concluded that injection locking leads to a loss of intermodal coherence. The electrical injection has an influence on the intensity spectrum only in the locking range (Fig. \ref{fig:fig3}f).

In real-life applications, QCL frequency combs have to withstand harsh conditions while maintaining coherence. Among these conditions is optical feedback. We illustrate the fatal effect of optical feedback on a free-running QCL frequency comb by replacing the attenuating polarizer (POL in Fig. \ref{fig:fig2}a) by a polished silicon wafer perpendicular to the QCL beam. In this configuration, the QCL is subject to both intense static feedback from the Si wafer as well as temporally varying feedback from the QWIP facet due to the scanning FTIR mirrors. While the electrical beatnote is narrow and stable if the beam is attenuated by the polarizer (Fig. \ref{fig:fig4}a), it becomes significantly broader and weaker upon exposure to strong optical feedback indicating the loss of coherence. This fatal effect of optical feedback is omnipresent in dual-comb spectrometers based on QCL combs. Expensive and bulky optical isolators have to be employed to ensure stable comb operation, impairing the capabilities of miniaturization.
In contrast, both the coherence and the phase characteristics of an injection locked comb are preserved even in presence of strong optical feedback (Fig. \ref{fig:fig4}c).
A prototype self-detected dual-comb setup highlights the enormous potential of coherent electrical injection locking for miniaturization. The light of two QCLs located on the same chip is shined directly into each other without any optically isolating elements in between (Fig. \ref{fig:fig4}c). When both lasers are locked, the self-detected dual-comb beat spectrum consists of numerous equidistant lines with a spacing of $\Delta f_{rep}=7.4\,$MHz. If the injection frequency of one laser is detuned by 200 kHz - thus leaving the locking range (Fig. \ref{fig:fig3}e) - the multiheterodyne signal becomes broad and no dual-comb lines are visible anymore. These results open up new avenues towards all-solid-state MIR spectrometers\cite{schwarz2014monolithically}, where stabilization of the combs and the ability to cope with intense feedback are vital. 

Our investigations demonstrate that electrical injection locking of MIR QCLs is a versatile technique that enables the generation of coherent frequency combs if the inherent spatio-temporal pattern of the electrical beatnote is taken into account. The fact that the repetition frequency is fixed by the injected signal can be utilized to stabilize the carrier envelope offset frequency against a narrow molecular absorption line via the driving current.
The possibility of all-electric stabilization using low-budget electronics, as those found in every mobile phone, will lead to a new class of miniaturized dual-comb spectrometers.

\section*{Methods}
\footnotesize
\setstretch{1.}
\noindent\textbf{Device}: The investigated QCL is uncoated and operating at 8$\:$\textmu m. The lase has a relatively low group delay dispersion as shown in supplementary Fig. 4. The laser is mounted epi-side-up on a copper submount. The temperature of the submount is kept at 15$^{\circ}$ for all measurements presented using a Peltier element and PTC5000 temperature controller. A HP 8341B synthesized sweeper is used for injection locking. The RF signal is injected close to the front end of the QCL cavity through 40 GHz two-terminal RF probes.
It has to be noted that due to the large parasitic capacitance, the RF signal is strongly damped 30-40dB. In case of an RF optimized device, split contacts would be required to prevent the simultaneous excitation of both ends with the same phase.

\vspace{0.1cm}

\noindent\textbf{SWIFTS}: The QWIP used to detect the optical beatnote is fabricated in square mesa geometry with 100 \textmu m side length. The mesa is connected to a coplanar transmission line with a short wirebond resulting in a cutoff frequency slightly below 10 GHz. The optical beatnote detected by the QWIP is amplified and mixed down to below 50 MHz. A Zurich Instruments HF2LI lock-in amplifier and the Helium-Neon trigger of the FTIR were used to record the SWIFTS interferograms. The complex sum of both quadrature interferograms can be written as
\begin{align}
(X + i Y)(\tau)=	\sum_n A_n A_{n-1}^* \biggl[ \cos\left(\frac{\omega_{r}\tau}{2}\right) +\cos \left(\omega_{n-\frac{1}{2}} \tau   \right)  \biggr].	\label{eq:SWIFTSinterferogram}	
\end{align}
Eq. \ref{eq:SWIFTSinterferogram} differs slightly from previously published work\cite{burghoff2015evaluating,singleton2018evidence} because our FTIR (Bruker Vertex 70v) moves both interferometer arms by $\pm \tau/2$ instead of just one arm by $\tau$.

\vspace{0.1cm}
\noindent\textbf{Self-detected dual-comb spectrum}: The two QCLs on the dual-comb chip are roughly 1$\,$mm apart from each other. The light emitted from the front facets of the dual-comb chip is collimated using an anti-reflection coated ZnSe lens with 1.5$\,$inch focal length. By aligning a mirror in front of the lens, the light of one laser is reflected into the other. The dual-comb beat around 8.7$\,$GHz is extracted directly from the laser using a RF probe and recorded with a spectrum analyzer (acquisition time$\,\approx0.2\,$s.)

\footnotesize

\providecommand{\noopsort}[1]{}\providecommand{\singleletter}[1]{#1}%

%\normalsize
%\bibliographystyle{naturemag}
%\bibliography{literature}
\providecommand{\noopsort}[1]{}\providecommand{\singleletter}[1]{#1}%

%************** Acknowledgement **************
\section*{Acknowledgements}
\footnotesize
\setstretch{1.}
\noindent This work was supported by the Austrian Science Fund (FWF) in the framework of "Building Solids for Function" (Project W1243), the projects "NanoPlas" (P28914-N27) and "NextLite" (F4909-N23). H.D. acknowledges the support by the ESF under the project CZ.\-02.2.69\-/0.0/0.0/16\-\_027/0008371. A.M.A was supported by the pro\-jects COM\-TERA - FFG 849614 and AFOSR FA9550-17-1-0340.

\section*{Author contributions}
\footnotesize
\setstretch{1.}
J.H. and B.S. built up the SWIFTS setup. J.H. carried out the experiments and wrote the manuscript with editorial input from B.S. and A.M.A.. H.D., A.M.A, and G.S. were responsible for MBE growth. B.S. developed the algorithm for the SWIFTS data processing and supervised this work. All authors contributed to analysing the results and commented on the paper.

\section*{Competing financial interests}
\footnotesize
\setstretch{1.}
The authors declare no competing financial interests.
\end{document}